\input amstex
\loadbold 
\documentstyle{amsppt}
\magnification=\magstep 1
\hsize29pc
\vsize42pc
\baselineskip=24truept
\def\Z{\Bbb Z}
\def\k{\bar{k}}
\def\z{\bar{z}}
\def\ta{\tilde{a}}
\def\tL{\tilde{L}}
\def\[{\left[}
\def\]{\right]}
\def\({\left(}
\def\){\right)}
\topmatter
\title  TRACE FORMULA   
FOR  A SYSTEM OF PARTICLES WITH ELLIPTIC POTENTIAL\endtitle
\author K.L.  Vaninsky\endauthor
\affil Department of Mathematics\\
Kansas State University\\
Manhattan, KS 66502
\endaffil
\email vaninsky@math.ias.edu\endemail
\thanks  The work is  supported by NSF grant DMS-9501002\endthanks
\abstract 
We consider classical particles on the line  with 
the  Weierstrass $\wp$ function as potential.  This system 
parameterizes special solutions of the KP equation.  We derive the trace 
formula which relates the Hamiltonian of the particle system 
to the residues of some Abelian differential 
(meromorphic one-form) on the spectral curve.  Such  formula is 
important for the construction action-angle variables  and
study  invariant Gibbs' states.
\endabstract
 
\endtopmatter
\rightheadtext{TRACE FORMULA FOR THE SYSTEM OF PARTICLES}
\document 
\subhead 1. Introduction\endsubhead
The  subject of this note is a system of $N$ classical particles on 
the line interacting with  the Hamiltonian
$$
H_N=\sum\limits_{n=1}^{N} {p_n^2\over 2} - 2 \sigma^2 \sum\limits_{n,m=1}^{N}
\wp(q_n-q_m).
$$
The parameter \footnote"*"{Should not be confused with the Weierstrass $\sigma$, 
see section 6.}  $\sigma=1$ corresponds to attractive particles and
$\sigma=i=\sqrt{-1}$ corresponds to repulsive particles.
The potential  is the Weierstrass  $\wp$ function with real period $2\omega$ 
and pure imaginary period $2\omega'$. The system includes the well known 
integrable potentials $\sin^{-2}x,$ $ \sinh^{-2}x$ and  $x^{-2}$; which correspond to
various degeneration of $\wp$. 
 
In a remarkable article Airault, McKean, Moser, \cite{AMM}, 
discovered a connection between  particles with rational or elliptic  
potential and the Korteweg-de Vries  equation
$$
u_t +{3\over 2} u u_x - {1\over 4} u_{xxx}=0.
$$
Few years later Krichever, \cite{K},  found an isomorphism  between 
particles with elliptic potentials and special solutions of 
the Kadomtzev-Petviashvilli (KP) equation
$$
{3\over 4} \sigma^2 u_{yy} = (u_t +{3\over 2} u u_x -{1\over 4} u_{xxx})_x.
$$
The case $\sigma=1$ corresponds to the KP-2 equation and $\sigma=i$ 
to KP-1. The KP equation arises as a compatibility condition for the zero 
curvature representation
$$
\[\sigma \partial_y-L_2,\partial_t -L_3\]=0,
$$
where
$$
\align
L_2&=\partial_x^2-u,\\
L_3&=\partial_x^3-{3\over2}u\partial_x-w.
\endalign
$$
Any such solution is associated with a spectral curve $\Gamma_N$ of genus $N$, 
defined by
$$
R_N(k,z)=\sum\limits_{n=0}^{N}r_n(z)k^n=\det (\tL +2k),
$$ 
where $\tL$ is a $N\times N$ matrix which depends on $q,p$ and $z$. 
The functions $r_n(z)$ are elliptic, so the curve $\Gamma_N$ is an $N$-sheeted covering 
of the elliptic torus. The matrix $\tL$ has a simple pole above $z=0$ and 
can be expanded in powers of $z$
$$
\tL= {1\over z} L^{(-1)} + O(1),
$$
where $L_{nm}=-2(1-\delta_{nm})$ is a constant matrix. This "zero order" 
approximation provides all the information needed to solve the direct 
spectral problem and obtain a formula for the solution in terms of Riemann 
theta functions, see \cite{K}.
 
In this note we address a different question. Is there a formula of the type 
$$
H_N=\sum\limits_{\alpha=2}^{N} I_{\alpha}',
$$
where $I_{\alpha}'$ are parameters of the Riemann surface associated with 
the system?  We give an affirmative answer to this question here. 
The formula is needed, see \cite{MCV1-2}, to express the canonical measure as 
$$
e^{-H} d\, vol= e^{-H} \prod dI\, d\phi = e^{-\sum I'} \prod dI\, d\phi,
$$
where $d\, vol$ is produced  from the basic symplectic 
structure\footnote"*"{Here $\omega$ is not the same as $2\omega$,  the period  of  the
elliptic functions.} $\omega=dp\wedge dq$ 
and $I$'s and $\phi$'s are classical action-angle variables constructed from $\omega$. 
After this is done one can try to  compute  the partition function,  \cite{V1}.

Now analysis of the direct spectral problem requires a "second order" 
approximation
$$
\tL= {1\over z} L^{(-1)} + L^{(0)} +zL^{(1)} + O(z^2),
$$
Such an approximation provides  the coefficients $k_1^{(0)}$ and $k_1^{(1)}$ 
(Theorem 9) for the expansion of the function $k$ 
$$
k_1(z)={N-1\over z} + k_1^{(0)} + z k_1^{(1)} + O(z^2)
$$
on the``upper'' sheet of the curve. The desired formula can be easily 
obtained by a simple 
application of Cauchy's theorem. Moreover, in the repulsive case the  parameters 
$I_{\alpha}'$ are real for 
all configurations of particles. In order to prove this we show that in the 
expansion
$$
k_{\alpha}(z)=-{1\over z} + k^{(0)}_{\alpha} +O(z),\quad \quad \alpha=2,\cdots, N.
$$
the coefficients $k^{(0)}_{\alpha}$ are distinct for all $\alpha=2,\cdots, N$ 
and a  generic configuration of particles. That  much  information can 
be obtained by perturbation techniques on  ``lower'' sheets.

Presumably $I_{\alpha}'$ are moduli of the corresponding $N$-sheeted covers and 
the actions relative to some  symplectic structure $\omega'$ 
on the phase space, but this is not proved. Compare \cite{V2} for the 
case of the cubic Schr\"{o}dinger curves. We will return to this issue elsewhere.

\subhead 2.  Elliptic solutions of KP hierarchy\endsubhead
Consider  the $N$ particle Hamiltonian on the line
$$
H_N=\sum\limits_{n=1}^{N} {p_n^2\over 2} - 2 \sigma^2 \sum\limits_{n,m=1}^{N}
\wp(q_n-q_m).
$$
The  Hamiltonian produces a system of first order equations of motion
$$
\align
\overset\bullet\to q_n & =\;\; {\partial H\over \partial p_n},  
\quad\quad\quad\quad\quad\quad\quad\quad\quad\quad \quad\quad\;\;
n=1,\hdots, N,\\
\overset\bullet\to p_n & =-{\partial H\over \partial q_n}=4 \sigma^2 
\sum\limits_{m\neq n} \wp'(q_n-q_m), \quad\quad n=1,\hdots, N.
\endalign
$$
The system can be written in the form
$$
\overset {\bullet\bullet}\to q_n= 4\sigma^2 \sum\limits_{m\neq n} \wp'(q_n-q_m),
\quad\quad n=1,\hdots, N. \tag 1
$$

The key step in the embedding of the particle system into  the 
elliptic solutions of the KP equation is the following theorem
\proclaim{Theorem 1}\cite{K}. The equations 
$$
\align
& \[\sigma \partial_y -\partial_x^2 + 2 \sum\limits_{n=1}^{N} \wp(x-q_n(y))\]
\psi =0, \\ 
\psi^{\dag} & \[\sigma \partial_y -\partial_x^2 + 2 \sum\limits_{n=1}^{n} 
\wp(x-q_n(y))\] =0  
\endalign
$$
have solutions of the form
$$
\align
\psi(x,y,k,z) &=\sum\limits_{n=1}^{N}a_n(y,k,z) \Phi(x-q_n,z) e^{kx+ \sigma^{-1}
k^2 y}\\
\psi^{\dag}(x,y,k,z) &=\sum\limits_{n=1}^{N}a^{\dag}_n(y,k,z) \Phi(-x+q_n,z) 
e^{-kx- \sigma^{-1} k^2 y},
\endalign
$$
where\footnote"*"{$\sigma(z)$ denotes the Weierstrass function.}  
$\Phi(x,z)={\sigma(z-x)\over \sigma(z)\sigma(x)}e^{\zeta(z) x}$, 
if and only if $q_n(y)$ satisfy the system of equations (1).
\endproclaim

The proof is obtained by requiring that singularities of the form  
$(x-q_n)^{-2}$ and $(x-q_n)^{-1}$ vanish. This condition can be written 
in a  compact  form with the aid of $N\times N$ matrices $L$ and $M$ 
$$
\align
L_{nm}& = \sigma p_n \delta_{nm} +2\Phi(q_n-q_m,z)(1-\delta_{nm}),\\
M_{nm}& = \(-\wp(z)+ 2\sum\limits_{s\neq n}\wp(q_n-q_s)\)\delta_{nm} 
+ 2 \Phi'(q_n-q_m,z) (1-\delta_{nm}). \endalign
$$

\proclaim{Lemma 2} \cite{K}. The vectors $a(y,k,z)$ and $a^{\dag}(y,k,z)$ 
satisfy the equations
$$
(L+ 2k) a=0 \quad \quad (\sigma\partial_y +M) a=0
$$
and 
$$
a^{\dag}(L+ 2k) =0 \quad \quad a^{\dag} (\sigma\partial_y +M) =0.
$$
\endproclaim

These equations determine the curve $\Gamma_N$ which is the subject of the 
next section.

\subhead 3. Riemann surface\endsubhead
The matrix $L$ can be simplified using a gauge transformation
$$
L= G \tL G^{-1},
$$
where $G_{nm}=e^{\zeta(z)q_n}\delta_{nm}$. Then 
$$
\tL_{nm}= \sigma p_n \delta_{nm} +2\Phi_0(q_n-q_m,z) (1-\delta_{nm})
$$
and
$$
\Phi_0(x,z)={\sigma(z-x)\over \sigma(z) \sigma(x)}.
$$ 
The existence 
of a nontrivial vector $a:\; (\tL+ 2k)a=0$ implies that $R_N(k,z)=\det (\tL +
2k)$ vanishes and this condition determines the curve.  
We denote by $P,Q,$ {\it etc.,}  points $(k,z)$ on the curve.
\proclaim{Lemma 3}\cite{K}. The determinant  $R_N(k,z)$ 
can be written in the form
$$
R_N(k,z)=\sum\limits_{n=0}^{N} r_n(z) k^n,
$$
where $r_n(z)$ are elliptic functions of $z$.
\endproclaim

The curve $\Gamma_N$ is an $N$ sheeted covering of the elliptic curve. The next 
lemma describes symmetries of the curve.

\proclaim{Lemma 4} $(i).\quad \sigma= 1$. The curve $\Gamma_N$ admits 
the antiholomorphic involution 
$$
\tau_1:\quad (k,z) \rightarrow (\bar{k}, \bar{z}).
$$
$(ii).\quad \sigma= i$. The curve $\Gamma_N$ admits 
the antiholomorphic involution 
$$
\tau_i:\quad (k,z) \rightarrow (-\bar{k},-\bar{z}).
$$
\endproclaim
\demo\nofrills{Proof.\usualspace} $(i)$. For rectangular lattice 
$\overline{\sigma(z)}= \sigma(\bar{z})$ and 
$\overline{\Phi_0(x,z)}=\Phi_0(\bar{x}, 
\bar{z})$. Therefore  $\overline{R_N(k,z)}=R_N(\bar{k},\bar{z})$.

$(ii)$. Note first $\Phi_0(x,-z)=-\Phi_0(-x,z)$ and $\overline{ip +2k}= 
-(ip +2(-\bar{k})).$ Therefore 
$\overline{R_N(k,z)}=(-1)^N R_N(-\bar{k},-\bar{z})$. 
The proof is finished. 
\qed
\enddemo

The function $\Phi_0(x,z)$ has a simple pole at $z=0$ and can be expanded in 
powers of $z$
$$
\Phi_0(x,z)= -{1\over z} +  \zeta(x) +{1\over 2} (\wp(x)- \zeta^2(x))z + O(z^2).
$$
Therefore\footnote"*"{We omit $\tilde{}$ above $L$ and $a$ to simplify 
the notations.},
$$
\tL(p,q,z)= {1\over z} L^{(-1)}(p,q) + L^{(0)}(p,q)+ z L^{(1)}(p,q) +O(z^2),
$$
where
$$
\align
L_{nm}^{(-1)} &= -2 (1-\delta_{nm}),\\
L_{nm}^{(0)} \, &= \sigma p_n \delta_{nm} + 2 \zeta(q_n-q_m)(1-\delta_{nm}),\\
L_{nm}^{(1)} \, &= (\wp -\zeta^2)(q_n-q_m)(1-\delta_{nm}).              
\endalign
$$
The matrix $L^{(-1)}$  is a   constant matrix. Its spectrum 
and eigenvectors can be easily computed. 

Let 
$$
a_{\alpha}=\(\matrix 
1 \\
e^{i\beta_{\alpha}}\\
\vdots\\
e^{i\beta_{\alpha}(N-1)}\endmatrix \)\quad\text{and} \quad
\beta_{\alpha}={2\pi\over N}(\alpha-1),\quad \alpha=1,\hdots ,N.
$$ 
Then
$$
\(L^{(-1)}+2k_{\alpha}\)a_{\alpha}=0
$$
with $k_{\alpha}=-1$ for $\alpha=2,\hdots, N$ and $k_1=N-1$. 

The necessary information about the curve is obtained by perturbation of this 
trivial case. The eigenvectors $a_{\alpha}(z)$ and eigenvalues $k_{\alpha}(z)$ 
can be expanded in  power series in $z$ 
$$
\align
a_{\alpha}(z)&= a_{\alpha}^{(0)} + a_{\alpha}^{(1)} z+  a_{\alpha}^{(2)}z^2 + 
\hdots, \quad \quad  a_{\alpha}^{(0)}= a_{\alpha},\\
k_{\alpha}(z)&={1\over z} k_{\alpha}^{(-1)} + k_{\alpha}^{(0)} + 
k_{\alpha}^{(1)}z+\hdots, \quad\quad  
k_{\alpha}^{(-1)}= k_{\alpha}. 
\endalign
$$
The index ``$\alpha$'' labels the sheets of the curve $\Gamma_N$.  We call 
the sheet $\alpha=1$ the ``upper'' sheet. In fact, the upper sheet is 
distinguished by $k^{(-1)}$, the leading term of the asymptotics. The ``lower'' 
sheets $(\alpha \geq 2)$ are distinguished for generic configuration of particles 
by  different  values of $k^{(0)}$ since all $k^{(-1)}=-1$. This is proved in  
Lemma 5. The proof  of  Lemma 6 shows that all leading terms $a_{\alpha}^{(0)}$   
are distinct. This implies that all ``lower'' sheets can be indexed  
according to these asymptotics  and $a_{\alpha}^{(0)} = a_{\alpha}$.

\proclaim{Lemma 5} For generic configuration of particles $k^{(0)}$ are distinct
\endproclaim
\demo\nofrills{Proof.\usualspace}To prove the statement  
we need first order perturbation theory for multiple eigenvalues.

We  choose $N-1$ vectors $e^{(0)}$ in the subspace generated by
$a^{(0)}_{\alpha},\; \alpha=2,\hdots, N$;
$$
e^{(0)} =\sum\limits_{\alpha=2}^{N}\eta_{\alpha} a^{(0)}_{\alpha},
$$
where the $\eta$'s depend on $e^{(0)} $ and are such that
$$
\(\tL(z)+ 2k(z)\) e_{\gamma}(z)=0.
$$
$L(z), \, k_{\gamma}(z),\, e_{\gamma}(z)$ can be expanded in
{\it integer}  powers of $z$
$$
\align
k_{\gamma}(z)&={1\over z} k_{\gamma}^{(-1)} +  k_{\gamma}^{(0)}  + z
k_{\gamma}^{(1)}+ \hdots, \\
e_{\gamma}(z)&= e^{(0)}_{\gamma} + z e^{(1)}_{\gamma} + O(z^2).
\endalign
$$
Now we collect terms in the identity
$$
\[{1\over z} L^{(-1)} + L^{(0)}  +\hdots +  {2\over z} k^{(-1)} + 2 k^{(0)} +
\hdots \] \[ e^{(0)}  +  z e^{(1)} + \hdots \]=0.
$$
Terms in $z^0$ produce
$$
L^{(-1)} e^{(1)} + L^{(0)} e^{(0)}  + 2 k^{(-1)} e^{(1)}+ 2 k^{(0)} e^{(0)}=0,
$$
and
$$
(L^{(-1)} e^{(1)}, a_{\alpha}^{(0)}) + (L^{(0)} e^{(0)}, a_{\alpha}^{(0)}) +
 2 k^{(-1)} (e^{(1)},a_{\alpha}^{(0)})+2 k^{(0)} (e^{(0)}, a_{\alpha}^{(0)})=0.
$$
Using the selfadjointness of $L^{(-1)}$
$$
(L^{(-1)} e^{(1)}, a_{\alpha}^{(0)})= (e^{(1)},L^{(-1)} a_{\alpha}^{(0)})=
- 2 k^{(-1)} (e^{(1)}, a_{\alpha}^{(0)}).
$$
Therefore
$$
(L^{(0)} e^{(0)}, a_{\alpha}^{(0)})= -2 k^{(0)}(e^{(0)}, a_{\alpha}^{(0)})
$$
or
$$
\sum\limits_{\alpha=2}^{N} \eta_{\alpha}(L^{(0)} a_{\alpha}^{(0)},
a_{\alpha'}^{(0)})= -2 k^{(0)} \eta_{\alpha'}.
$$
The eigenvalues  $-2 k^{(0)}$ of the matrix
$(L^{(0)} a_{\alpha}^{(0)},a_{\alpha'}^{(0)}),\; \alpha, \alpha'= 2,\hdots,N;$
are distinct for   generic configuration of particles. 
\qed
\enddemo
\proclaim{Lemma 6} For all configurations of particles the ``zero'' order approximations
$$
\ta^{(0)}=\(\matrix
 \ta^{(0)}(1)\\
\vdots\\
\ta^{(0)}(N)\endmatrix \)
$$
of the eigenvectors
$\ta_{\alpha}(z)= \ta_{\alpha}^{(0)} + \ta_{\alpha}^{(1)} z+  \hdots$  of the
spectral problem $(\tL+ 2k) \ta=0$ normalized by the
condition $\ta^{(0)} (1)=1$ are  given by  the formula
$$
\ta_{\alpha}^{(0)}=\(\matrix
1 \\
e^{i\beta_{\alpha}}\\
\vdots\\
e^{i\beta_{\alpha}(N-1)}\endmatrix \)\quad\text{and} \quad
\beta_{\alpha}={2\pi\over N}(\alpha-1),\quad \alpha=1,\hdots ,N.
$$
\endproclaim
\demo\nofrills{Proof.\usualspace} If ${\Cal A}$ as $n\times n$ matrix then
${\Cal A}{\Cal A}^{\wedge}= \det {\Cal A} \,I$, where ${\Cal A}^{\wedge}$ is the
matrix which consists of auxiliary minors of ${\Cal A}$, see also \cite{KNS}.  For 
any column $r=1, \hdots ,N$
$$
\ta(p)={\[\tL +2k\]^{\wedge}_{pr}\over \[\tL +2k\]^{\wedge}_{1r}} 
$$
in the vicinity of $z=0$\footnote"*"{ Note that $ \ta$ is not a function on the
curve, since the entries of the matrix $\tL +2k$ are not elliptic functions.}.

We know that $k(z)=-{1\over z} + k^{(0)} + \dots $ and
$$
\tL=-{2\over z}[E -I] +\tL^{(0)}  +\hdots , \quad\quad E_{nm}=1.
$$
Then
$$
\[\tL +2k\]^{\wedge}_{pr}= \[ -{2E\over z} + [\tL^{(0)} + 2 k^{(0)} ] + \hdots \]^{\wedge}
_{pr}.
$$
Since $\text{rank}\,  E =1$ we have
$$
\[\tL +2k\]^{\wedge}_{pr}= -{2\over z} \sum\limits_{s=1}^{N}  [\tL^{(0)}
+ 2 k^{(0)}]^{\wedge}_{\hat{s}\; pr} + \hdots,
$$
where the subscript $\hat{s}$ means that the $s$-th column is replaced by
$\(\matrix
 1\\
\vdots\\
1\endmatrix \).$

For a generic configuration $k^{(0)}$ are distinct on all sheets of the curve. This
and the formula for $\ta(p)$  imply that  all $\ta^{(0)}$  are also distinct and 
therefore match  $\ta_{\alpha}^{(0)}$.
Since all $\ta_{\alpha}^{(0)}$ are fixed  the statement is true for all
configurations of particles.
\qed
\enddemo

The following two lemmas are simple consequences of the discussion above
\proclaim{Lemma 7}\cite{K}. The determinant $R_N(k,z)$ can 
be written in the form
$$
R_N(k,z)= 2^N\(k-({N-1\over z}+k_1^{(0)} + \hdots )\) 
\prod\limits_{\alpha=2}^{N} \(k-(-{1\over z} + k_{\alpha}^{(0)} +\hdots)\).
$$
\endproclaim

The genus of the curve $\Gamma_N$ can be easily computed using the Riemann-Hurwitz 
formula.

\proclaim{Lemma 8}\cite{K, KBBT}. $(i)$ The elliptic case: $2\omega, 2\omega' 
< \infty$. For generic configuration of particles the genus of the 
curve $\Gamma_N$ is  $N$. 

$(ii)$ The rational case: $2\omega, 2\omega' = \infty$. The genus of the curve 
$\Gamma_N$ is  $0$. 
\endproclaim

\subhead 4. Asymptotics for $k(z)$\endsubhead
The main result of this section is the following
\proclaim{Theorem 9} On the "upper" sheet for $k_1(z)$ the following 
asymptotics hold
$$
k_1(z)= {N-1\over z} -{\sigma P_N\over 2N} + z\({\sigma^2 H_N\over 2N^2}
-{\sigma^2 P_N^2\over 4N^3}\) + O(z^2),
$$
where
$$
H_N=K_N-\sigma^2 V_N= \sum\limits_{n=1}^{N} {p_n^2\over 2} -
2 \sigma^2 \sum\limits_{n,m=1}^{N} \wp(q_n-q_m),
$$
and
$$
P_N= \sum\limits_{n=1}^{N} p_n.
$$
\endproclaim

To prove the theorem we need the following lemma which is the second order 
perturbation theory of simple eigenvalues adapted to our considerations.

\proclaim{Lemma 10} The following identities hold

$$
\align
2k^{(0)} & =-(L^{(0)}a^{(0)},a^{(0)}), \\
2k^{(1)}& ={1\over 2N}\sum\limits_{\alpha=2}^{N} (L^{(0)} a^{(0)}_{\alpha}, 
a^{(0)}) \times (L^{(0)} a^{(0)}, a^{(0)}_{\alpha}) - 
(L^{(1)} a^{(0)}, a^{(0)}),
\endalign
$$
where 
$$
(f,g)={1\over N}\sum\limits_{n=1}^{N} f_n\bar{g}_n,
$$
for any $f,g \in {\Bbb C}^N$.
\endproclaim 
\demo\nofrills{Proof.\usualspace}
We  start with the identity
$$
\align
({1\over z} L^{(-1)} + & L^{(-0)} + zL^{(1)} + \hdots \\
&+ {2k^{(-1)}\over z} 
+2 k^{(0)} +2 k^{(1)} +\hdots) (a^{(0)} + z a^{(1)} +\hdots)=0.
\endalign
$$
Collecting the terms in $z^{-1}$
$$
\(L^{(-1)} + 2k^{(-1)} \) a^{(0)}=0.
$$
Note, 
$$
a^{(0)} = a^{(0)}_1=\(\matrix
1 \\
1\\
\vdots\\
1\endmatrix \).
$$
Therefore $k^{(-1)} = N-1$, as it has to be.

Now we collect terms with $z^0$
$$
L^{(-1)}a^{(1)} + L^{(0)}a^{(0)} + 2 k^{(-1)} a^{(1)} +2 k^{(0)} a^{(0)}=0
\tag 2
$$
and
$$
(L^{(-1)}a^{(1)} , a^{(0)}) + (L^{(0)}a^{(0)}, a^{(0)}) + 
(2 k^{(-1)} a^{(1)},a^{(0)}) + (2 k^{(0)} a^{(0)}, a^{(0)})= 0. \tag 3
$$
Note
$$
(L^{(-1)}a^{(1)} , a^{(0)})= (a^{(1)} ,L^{(-1)} a^{(0)})= (a^{(1)}, 
-2k^{(-1)} a^{(0)}), \tag 4
$$
which implies that the two terms in (3) cancel each other. From the definition 
$(a^{(0)},a^{(0)}) =1$ we obtain the first statement of the lemma. 

Now we derive the formulas for $a^{(1)}$ on the "upper" sheet, which will be 
useful later on.
Multiply (2) by $a^{(0)}_{\alpha}, \; \alpha=2,\hdots, N$,
$$
(L^{(-1)}a^{(1)} , a^{(0)}_{\alpha}) + (L^{(0)}a^{(0)}, a^{(0)}_{\alpha}) +
(2 k^{(-1)} a^{(1)},a^{(0)}_{\alpha}) + 
(2 k^{(0)} a^{(0)}, a^{(0)}_{\alpha})= 0. \tag 5
$$
The last term vanishes due to the orthogonality of eigenvectors 
corresponding different eigenvalues. Similar to $(4)$
$$
(L^{(-1)}a^{(1)} , a^{(0)}_{\alpha})= (a^{(1)}, L^{(-1)} a^{(0)}_{\alpha}) =
2( a^{(1)} , a^{(0)}_{\alpha})
$$
due to 
$$
\( {1\over z} L^{(-1)} + 2 ( -{1\over z}) \) a^{(0)}_{\alpha}= 0, 
\quad \quad \alpha=2, \hdots, N.
$$
Therefore
$$
2(a^{(1)},a^{(0)}_{\alpha}) + (L^{(0)}a^{(0)}, a^{(0)}_{\alpha}) + 2(N-1) 
(a^{(1)}, a^{(0)}_{\alpha})=0
$$
and
$$
(a^{(1)},a^{(0)}_{\alpha}) = -{1\over 2N} (L^{(0)}a^{(0)},  a^{(0)}_{\alpha}),
\quad \quad \alpha=2,\hdots, N.
$$
The condition $||a(z)||^2=1+O(z^2)$ implies
$$
\align
1=(a(z),a(z)) &=(a^{(0)} + z a^{(1)} +\hdots, a^{(0)} + z a^{(1)} +\hdots) \\
&=(a^{(0)},a^{(0)}) + z\[(a^{(0)},  a^{(1)}) + (a^{(1)}, a^{(0)})\] + \hdots.
\endalign
$$
and $ (a^{(1)} , a^{(0)})=0$. Therefore
$$
a^{(1)}=\sum\limits_{\alpha=2}^{N} a_{\alpha}^{(0)} ( a^{(1)},
a^{(0)}_{\alpha})= -{1\over 2N} \sum\limits_{\alpha=2}^{N} a_{\alpha}^{(0)}
(L^{(0)} a^{(0)},  a_{\alpha}^{(0)}).
$$

In order to prove the second formula of the lemma we collect terms with $z^1$
$$
L^{(-1)}a^{(2)} + L^{(0)}a^{(1)} + L^{(1)}a^{(0)} + 2k ^{(-1)} a^{(2)} 
+ 2k^{(0)}a^{(1)} + 2k^{(1)}a^{(0)} =0
$$
and 
$$
\align
(L^{(-1)}a^{(2)}, a^{(0)}) &+ (L^{(0)}a^{(1)}, a^{(0)}) + 
(L^{(1)}a^{(0)}, a^{(0)})  +   \\
&+ 2k ^{(-1)} ( a^{(2)} , a^{(0)})
 +  2k^{(0)} ( a^{(1)} , a^{(0)}) + 2k^{(1)} ( a^{(0)}, a^{(0)}) =0. \tag 6
\endalign 
$$
Note 
$$
(L^{(-1)}a^{(2)}, a^{(0)})= ( a^{(2)}, L^{(-1)}  a^{(0)}) = -2k ^{(-1)}
(a^{(2)}, a^{(0)})
$$
and two of the terms in the formula (6) vanish. Using the normalization conditions
\break 
$(a^{(0)}, a^{(0)})=1$ and $(a^{(1)}, a^{(0)})=0$ we obtain
$$
2k^{(1)} = -(L^{(0)} a^{(1)},  a^{(0)}) - (L^{(1)} a^{(0)},  a^{(0)}).
$$
Using the formula for $a^{(1)}$, we finally have
$$
2k^{(1)} ={1\over 2N}\sum\limits_{\alpha=2}^{N} (L^{(0)} a^{(0)}_{\alpha},
a^{(0)}) \times (L^{(0)} a^{(0)}, a^{(0)}_{\alpha}) -
(L^{(1)} a^{(0)}, a^{(0)}).
$$
The lemma is proved.
\qed
\enddemo 

Now we use  the lemma to compute the coefficients $k^{(0)}$ and 
$k^{(1)}$. Note, first, that $L^{(0)}$ can be split into two parts
$L^{(0)}= \sigma A + B$, where 
$$
A_{mn}= p_n \delta_{nm}\quad \text{and}\quad \quad B_{nm}= 
2\zeta(q_n- q_m) (1- \delta_{nm}).
$$ 
The matrix $A$ is symmetric and  $B$ is skew-symmetric 
$$
\align
2k^{(0)}&= - ( L^{(0)}a^{(0)}, a^{(0)}) = -\sigma (A a^{(0)}, a^{(0)}) - 
(B a^{(0)}, a^{(0)}) \\
&= -{\sigma P_N\over N}.
\endalign
$$
The term with $B$ vanishes due to skew-symmetry.  

The computation  of the next term $k^{(1)}$ is much more involved. 
Using the decomposition for $L^{(0)}$ we have
$$
\align
2k^{(1)}= {1\over 2N} & \sum\limits_{\alpha=2}^{N} \[\sigma (A 
a_{\alpha}^{(0)}, a^{(0)}) + ( B a_{\alpha}^{(0)}, a^{(0)})\] \times
\[\sigma (A 
a^{(0)}, a^{(0)}_{\alpha}) +( B a^{(0)}, a^{(0)}_{\alpha}) \]\\
- &(L^{(1)}a^{(0)}, a^{(0)}) = I + II.
\endalign
$$
\subsubhead Step 1 \endsubsubhead
 Our goal is to evaluate the first term $I$. Arguments for the attractive and 
the repulsive case are different and we consider these two cases separately.

\noindent{\bf Repulsive case. } $\sigma=i$. Using the symmetry of $A$ and 
skew-symmetry of $B$, 
$$
\align
{1\over 2N} & \sum\limits_{\alpha=2}^{N} \[i( A
a_{\alpha}^{(0)}, a^{(0)}) + ( B a_{\alpha}^{(0)}, a^{(0)})\] \times
\[i( A
a^{(0)}, a^{(0)}_{\alpha}) +( B a^{(0)}, a^{(0)}_{\alpha}) \]\\
=-{1\over 2N} & \sum\limits_{\alpha=2}^{N} | i( A
a^{(0)}, a^{(0)}_{\alpha}) +( B a^{(0)}, a^{(0)}_{\alpha})|^2\\
=-{1\over 2N} & \sum\limits_{\alpha=1}^{N} | i( A
a^{(0)}, a^{(0)}_{\alpha}) +( B a^{(0)}, a^{(0)}_{\alpha})|^2 + 
{1\over 2N}   |( Aa^{(0)}, a^{(0)})|^2.
\endalign
$$
The last term can be easily estimated
$$
{1\over 2N} |( Aa^{(0)}, a^{(0)})|^2 = {1\over 2N^3} P_N^2.
$$
To estimate the first term, we introduce $\alpha'$ such that
$$\alpha'=N-\alpha +2.
$$
Then 
$$
\beta_{\alpha'}={2\pi \over N}(\alpha'-1)= {2\pi \over N}(N-\alpha +1)
$$
and
$$
\beta_{\alpha}+ \beta_{\alpha'}  \equiv 0 \quad (\text{mod}\; 2\pi).
$$
Now 
$$
\align
-{1\over 2N} & \sum\limits_{\alpha=1}^{N} 
| i( A a^{(0)}, a^{(0)}_{\alpha}) +( B a^{(0)}, a^{(0)}_{\alpha})|^2\\
=-{1\over 4N} & \sum\limits_{\alpha=1}^{N} | 
i( A a^{(0)}, a^{(0)}_{\alpha}) +( B a^{(0)}, a^{(0)}_{\alpha})|^2 + 
|\overline{i( A
a^{(0)}, a^{(0)}_{\alpha'}) +( B a^{(0)}, a^{(0)}_{\alpha'})}|^2\\
=-{1\over 4N} & \sum\limits_{\alpha=1}^{N} 
| i( A a^{(0)}, a^{(0)}_{\alpha}) +( B a^{(0)}, a^{(0)}_{\alpha})|^2 + 
|-i( A a^{(0)}, a^{(0)}_{\alpha}) +( B a^{(0)}, a^{(0)}_{\alpha})|^2.
\endalign
$$ 
Using the identity $|a+b|^2 + |a-b|^2 = 2 |a|^2 + 2|b|^2$ we obtain
$$
I= -{1\over 2N}\sum\limits_{\alpha=1}^{N} |(A a^{(0)}, a^{(0)}_{\alpha})|^2 + 
 |(B a^{(0)}, a^{(0)}_{\alpha})|^2 + {1\over 2N^3} P_N^2.
$$

\noindent{\bf Atractive case. } $\sigma=1$. Again using properties of $A$ and
$B$ we have
$$
\align
{1\over 2N} & \sum\limits_{\alpha=2}^{N} \[(A
a_{\alpha}^{(0)}, a^{(0)}) + ( B a_{\alpha}^{(0)}, a^{(0)})\] \times
\[( A
a^{(0)}, a^{(0)}_{\alpha}) +( B a^{(0)}, a^{(0)}_{\alpha}) \]\\
= {1\over 2N} & \sum\limits_{\alpha=2}^{N} \[\overline{( A
a^{(0)}, a^{(0)}_{\alpha})}  -\overline{( B a^{(0)}, a^{(0)}_{\alpha})}\] 
\times \[(A a^{(0)}, a^{(0)}_{\alpha}) +( B a^{(0)}, a^{(0)}_{\alpha}) \]\\
={1\over 2N} & \sum\limits_{\alpha=2}^{N} | ( Aa^{(0)}, 
a^{(0)}_{\alpha})|^2 - | ( B a^{(0)},a^{(0)}_{\alpha})|^2 \\
&\phantom {kkkkkkkkkkk}+ 2i {1\over 2N} \Im  \sum\limits_{\alpha=2}^{N} 
\overline{( Aa^{(0)},a^{(0)}_{\alpha})} 
( B a^{(0)},a^{(0)}_{\alpha}). 
\endalign
$$
The second sum vanishes. Indeed,
$$
\align
2 \Im  &\sum\limits_{\alpha=2}^{N} 
\overline{( Aa^{(0)},a^{(0)}_{\alpha})} 
( B a^{(0)},a^{(0)}_{\alpha})\\
=&\Im  \sum\limits_{\alpha=1}^{N}
\overline{( Aa^{(0)},a^{(0)}_{\alpha})}
( B a^{(0)},a^{(0)}_{\alpha}) + 
\overline{(Aa^{(0)},a^{(0)}_{\alpha'})}( B a^{(0)},a^{(0)}_{\alpha'})\\
= &\Im \sum\limits_{\alpha=1}^{N}
\overline{( Aa^{(0)},a^{(0)}_{\alpha})} ( B a^{(0)},a^{(0)}_{\alpha}) + 
( Aa^{(0)},a^{(0)}_{\alpha}) \overline{( B a^{(0)},a^{( 0)}_{\alpha})}\\
=& 0.
\endalign
$$
Therefore, 
$$
I= -{1\over 2N} \sum\limits^{N}_{\alpha=1} - 
|( Aa^{(0)},a^{(0)}_{\alpha})|^2 + 
|( B a^{(0)},a^{(0)}_{\alpha})|^2 -{1\over 2N^3} P_N^2.
$$
Combining the results for two cases $\sigma^2=\pm 1$, we obtain 
$$
I= -{1\over 2N} \sum\limits^{N}_{\alpha=1} - \sigma^2
|( Aa^{(0)},a^{(0)}_{\alpha})|^2 + 
|( B a^{(0)},a^{(0)}_{\alpha})|^2 -{\sigma^2\over 2N^3} P_N^2.
$$
\subsubhead Step 2 \endsubsubhead Now we estimate $ \sum\limits^{N}_{\alpha=1} 
|( Aa^{(0)},a^{(0)}_{\alpha})|^2 $ and 
$|( B a^{(0)},a^{(0)}_{\alpha})|^2$. To do this we introduce for 
two polynomials 
$$
{\Cal P}(z)= \sum\limits_{k=1}^{N}p_k z^k, \quad \quad \quad 
{\Cal Q}(z)= \sum\limits_{k=1}^{N}q_k z^k,
$$
the inner product
$$
<{\Cal P}, {\Cal Q}> \equiv \sum\limits_{\alpha=1}^{N} {\Cal P}(e^{i 
\beta_\alpha}) \overline{{\Cal Q}(e^{i \beta_\alpha})}.
$$
It is easy to see that
$$
<{\Cal P}, {\Cal Q}> = N \sum\limits_{k=1}^{N} p_k \overline{q_k}.
$$
Therefore
$$
\align
\sum\limits^{N}_{\alpha=1}&  |( Aa^{(0)},a^{(0)}_{\alpha})|^2 = 
\sum\limits^{N}_{\alpha=1} | {1\over N}\sum\limits^{N}_{n=1} 
p_n e^{-i\beta_{\alpha}(n-1)}|^2\\
&={1\over N^2} < \sum\limits^{N}_{n=1} p_n z^n, \sum\limits^{N}_{n=1} p_n z^n> 
= {1\over N} \sum\limits^{N}_{n=1} p_n^2.
\endalign
$$
Similarly,
$$
\sum\limits^{N}_{\alpha=1}  |( Ba^{(0)},a^{(0)}_{\alpha})|^2= 
{1\over N} \sum\limits^{N}_{n=1} (b_n)^2, \quad \quad 
b_n= \sum\limits^{N}_{s=1} b_{ns}.
$$
Finally,
$$
\align
I&= {\sigma^2\over 2 N^2} \sum\limits^{N}_{n=1} p_n^2 - 
{1\over 2N^2}  \sum\limits^{N}_{n=1} b_n^2 - {\sigma^2\over 2N^3}P_N^2\\
& = {\sigma^2\over N^2}K_N - {1\over 2N^2} \sum\limits^{N}_{n=1} b_n^2 - 
{\sigma^2\over 2 N^3}  P_N^2.
\endalign
$$
\subsubhead Step 3\endsubsubhead Now using the expression for $II$\footnote"*"{In 
the rational case $\wp - \zeta^2\equiv 0$; $L^{(1)}$ and $II$ vanish 
identically.}
$$
II= -{1\over N} \sum\limits_{s\neq s'} (\wp - \zeta^2) (q_s - q_{s'}),
$$
we will estimate
$$
\align
I + II = &{\sigma^2\over N^2} K_N - {1\over 2N^2} \sum\limits_{n=1}^{N} 
\(\sum\limits_{s=1}^{N} 2\zeta(q_n-q_s)\)^2  -{\sigma^2\over 2N^3} P_N^2\\
&- {1\over N} \sum\limits_{s\neq s'}\wp(q_s-q_{s'}) + {1\over N} 
\sum\limits_{s\neq s'} \zeta^2(q_s-q_{s'}).
\endalign
$$
We will prove that
$$
\align
- {1\over 2N^2} \sum\limits_{n=1}^{N}& \(\sum\limits_{s=1}^{N} 
2\zeta(q_n- q_s)\)^2 + {1\over N}\sum\limits_{s\neq s'} \zeta^2(q_s-q_{s'})\\
& = {N-2\over N^2} \sum\limits_{s\neq s'}\wp(q_s-q_{s'}). \tag 7 
\endalign
$$
This would imply 
$$
I + II= {\sigma^2\over N^2} K_N -{1\over N^2}V_N - {\sigma^2\over 2N^3} P_N^2
$$
and complete the proof of the theorem.

\subsubhead Step 4 \endsubsubhead 
To prove the identity (7) we make some manipulations with the 
sums
$$
\align
- {2\over N^2}& \sum\limits_{n=1}^{N}\[\sum\limits_{s=1}^{N}\zeta(q_n- q_s)\]^2
+ {1\over N} \sum\limits_{s\neq s'} \zeta^2 ( q_s -q_{s'})\\
=& - {2\over N^2}\sum\limits_{s\neq n,\, s'\neq n}\zeta(q_n- q_s)\zeta(q_n- 
q_{s'})
+ {1\over N} \sum\limits_{s\neq s'} \zeta^2 ( q_s -q_{s'})\\
=& {1\over N^2} \[(N-2) \sum\limits_{s\neq s'} \zeta^2 ( q_s -q_{s'}) -
2 \sum\limits_{s\neq s'\neq n} \zeta(q_n-q_s) \zeta(q_n-q_{s'})\]\\
=& {1\over N^2} \[\sum\limits_{s\neq s'\neq n} \zeta^2(q_s-q_{s'}) + 
2\zeta(q_s-q_n) \zeta(q_n- q_{s'})\].  
\endalign 
$$
Fix  $k_1 < k_2 < k_3$ and consider all their permutations. Collect 
the terms with such indices
$$
\align
{2\over N^2} \sum\limits_{k_1 < k_2 < k_3} &\zeta^2(q_{k_1}- q_{k_2}) +
\zeta^2(q_{k_2}- q_{k_3}) + \zeta^2(q_{k_3}- q_{k_1}) \\
+ &2 \zeta(q_{k_1}-q_{k_3} ) \zeta(q_{k_3}-q_{k_2} ) 
+ 2 \zeta(q_{k_1}-q_{k_2} ) \zeta(q_{k_2}-q_{k_3} )\\
+ &2 \zeta(q_{k_2}-q_{k_1} ) \zeta(q_{k_1}-q_{k_3} )\\
={2\over N^2}  \sum\limits_{k_1 < k_2 < k_3} &\( \zeta(q_{k_1}- q_{k_2}) + 
 \zeta(q_{k_2}- q_{k_3}) +  \zeta(q_{k_3}- q_{k_1})\)^2.
\endalign
$$
Using the identity 
$$
\(\zeta(u) + \zeta(v) + \zeta(s)\)^2 = \wp(u) +  \wp(v) +  \wp(s)
$$
for $u+ v + s = 0$, we obtain
$$
{2\over N^2} \sum\limits_{k_1 < k_2 < k_3}  \wp(q_{k_1}- q_{k_2}) + 
\wp(q_{k_2}- q_{k_3}) +  \wp(q_{k_3}- q_{k_1}).
$$
The last expression is invariant under permutations and therefore it is 
equal
$$
\align
{2\over 6N^2} &\sum\limits_{s\neq s'\neq n} \wp(q_{s}- q_{s'}) + 
\wp(q_{s}- q_{n}) +  \wp(q_{s'}- q_{n})\\
= & {N-2\over N^2} \sum\limits_{s\neq s'} \wp(q_s-q_{s'}).
\endalign
$$
This completes the proof of the identity (7) and the theorem.

{\bf Example.} For $N=2$ the equation defining the curve $\Gamma_2$ has 
the form
$$
k^2 +k {\sigma P\over 2} +\({\sigma^2 P^2\over 8} - {\sigma^2 H\over 4} -
\wp(z)\)=0.
$$
Solving the quadratic equation 
$$
k_1(z)= -{\sigma P\over 4} +{1\over 2}\sqrt{{\sigma^2 P^2\over 4} - 
4\({\sigma^2 P^2\over 8} - {\sigma^2 H\over 4} - \wp(z)\)}.
$$
Expanding $k_1(z)$ at $z=0$ we obtain
$$
k_1(z)= {1\over z} - {\sigma P\over 4} +z \( {\sigma^2 H\over 8} -
{\sigma^2 P^2\over 32}\) +O(z^2).
$$

\subhead  5. Trace formula\endsubhead 
If the total momentum vanishes, $P_N=0$, then the result of  Theorem 9
becomes
$$
k_1(z)= {N-1\over z} +z {\sigma^2 H_N \over 2N^2}  + O(z^2). \tag 8 
$$
\proclaim{Theorem 11} The following identity holds 
$$
{\sigma^2\over 2N^2} H_N= \sum\limits_{\alpha=2}^{N} I_{\alpha}',
$$
where $I_{\alpha}'= -k^{(1)}_{\alpha}$. 

$\sigma =i.$ The antiinvolution $\tau_i$ does not permute the sheets of the curve. 
The variables $I_{\alpha}'$ are real for all configurations of particles.

$\sigma=1.$ For some configurations of particles the antiinvolution $\tau_1$ 
permutes some lower sheets  $\alpha$ and $\tau_1 \alpha$. It also 
leaves the other sheets   invariant. Corresponding variables $I'_{\alpha}$ and 
$I'_{\tau_1 \alpha}$ form complex conjugate paires. All other $I'$s corresponding 
invariant sheets are real.
\endproclaim
\demo\nofrills{Proof.\usualspace} By Cauchy's theorem
$$
{1\over 2\pi i} \int_{\gamma_1} k(P) \wp(z(P)) dz(P)= - 
\sum\limits_{\alpha=2}^{N}{1\over 2\pi i} \int_{\gamma_{\alpha}} 
k(P) \wp(z(P)) dz(P),
$$
where $\gamma_{\alpha}$ is a small contour surrounding $P_{\alpha}$, the point 
on the $\alpha$'th sheet above $z=0$. The asymptotics (8) implies the result. 

The second part of the proof  is different for repulsive and attractive 
cases. We treat them separately.

\noindent{\bf Repulsive case. } $\sigma=i$. 
If the pair $(k,z)$ satisfies 
$$
k(z)={1\over z} k_{\alpha}^{(-1)} + k_{\alpha}^{(0)} +
k_{\alpha}^{(1)}z+\hdots
$$
in the vicinity of $P_{\alpha}$, then the pair $(-\k,-\z)$ also satisfies 
similar expression in the vicinity of $P_{\tau_i \alpha}$ 
$$
-\k(z)={1\over -\z} k_{\tau_i \alpha}^{(-1)} + k_{\tau_i \alpha}^{(0)} +
k_{\tau_i \alpha}^{(1)}(-\z) +\hdots.
$$
Therefore 
$$
k(z)={1\over z} \bar{k}_{\tau_i \alpha}^{(-1)} - 
\bar{k}_{\tau_i \alpha}^{(0)} + \bar{k}_{\tau_i \alpha}^{(1)}z-\hdots.
$$
Comparision shows that $k^{(n)}_{\alpha}=\bar{k}_{\tau_i \alpha}^{(n)}$  for $n$ odd and 
$k^{(n)}_{\alpha}=- \bar{k}_{\tau_i \alpha}^{(n)}$  for $n$ even. 

We know that $k_1^{(-1)}=N-1$ and $k_{\alpha}^{(-1)}=-1$ for $\alpha=2,\hdots, N$ so that 
$\tau_i$ leaves the upper sheet invariant.

The Lemma 5 states  that the values of $k^{(0)}_{\alpha}$ 
are distinct on different sheets of the 
curve for a generic configuration of perticles. 
The matrix $(L^{(0)} a_{\alpha}^{(0)},a_{\alpha'}^{(0)})$ 
(see the proof of Lemma 5) 
is skew symmetric and therefore all $k_{\alpha}^{(0)}$ are pure imaginary: 
$\bar{k}_{\alpha}^{(0)}=-k_{\alpha}^{(0)}$. 
These together with the identity $\bar{k}_{\alpha}^{(0)}=-k_{\tau_i \alpha}^{(0)}$ imply that $\tau_i \alpha= \alpha$ 
and antiinvolution $\tau_i$ does not permute sheets. Therefore $k_{\alpha}^{(1)}=
\bar{k}_{\alpha}^{(1)}$ and approximation arguments 
complete the proof.

\noindent{\bf Attractive case. } $\sigma=1$. Arguments as before lead to 
$k_{\alpha}^{(n)}=\bar{k}_{\tau_1 \alpha}^{(n)}$ for all $n$. Skew-symmetry of 
the matrix $L^{(0)}$ is
lost but for generic configuration of particles $k_{\alpha}^{(0)}$ are distinct. 
The antiinvolution $\tau_1$ leaves the upper sheet invariant, but is can permute the lower sheets of the curve. An example after the proof of the Theorem shows how it happens 
for $N=3$. Therefore, some lower sheets $\alpha$ and 
$\tau_1 \alpha$ are interchanged by antiinvolution $\tau_1$ and 
some  are invariant. These and $k_{\alpha}^{(-1)}=
\bar{k}^{(-1)}_{\tau_1 \alpha}$, together with approximation arguments complete 
the proof of  Theorem.
\qed
\enddemo

{\bf Example.} For $N=3$ we can explicitly compute expansion for $k_{\alpha}
(z)$ at $z=0$. In this case
$$
R_3(k,z)= 8k^3 + 4\sigma P k^2 + \[ \sigma^2 ( P^2 -2H) -24 \wp(z)\] k
+ 8\sigma  J + 8\wp'(z),
$$
where
$$
8J= \sigma^2 p_1 p_2 p_3 + 4 p_1 \wp(q_2-q_3) + 
4 p_2 \wp(q_1-q_3)+ 4 p_3 \wp(q_1-q_2).
$$
If $P=0$  the equation of the curve $\Gamma_3$ becomes
$$
k^3-k\[{\sigma^2 H\over 4} + 3 \wp(z)\] + \sigma J +\wp'(z)=0.
$$
Cardano's formula for the roots
$$
k^3 +pk -q=0
$$
has the form
$$
\align
k_1(z)&=\quad \root3\of{{q\over2} +\sqrt{R}} + \quad 
\root3\of{{q\over2} -\sqrt{R}},\\
k_2(z)&=\omega^2 \root3\of{{q\over2} +\sqrt{R}} + 
\omega \root3\of{{q\over2} -\sqrt{R}},\\
k_3(z)&=\omega \root3\of{{q\over2} +\sqrt{R}} + 
\omega^2  \root3\of{{q\over2} -\sqrt{R}},
\endalign
$$
where $\omega=e^{2\pi i\over 3},\; R={q^2\over 4} + {p^3\over 27}.$
After elementary but tiresome calculations we obtain
$$
\align
k_1(z)&=\;\;{2\over z} + z{\sigma^2 H\over 18}  + O(z^2),\\
k_2(z)&=-{1\over z} + {\sigma \sqrt{3H}\over 6} + z \(\;{J \over 
 \sqrt{3H} } -{\sigma^2 H\over 36}\) +O(z^2),\\
k_3(z)&=-{1\over z} - {\sigma \sqrt{3H}\over 6} + z \(-{J \over 
\sqrt{3H} } -{\sigma^2 H\over 36}\) +O(z^2).
\endalign
$$

Consider the case $\sigma=1$. If $H< 0$, then $\tau_1$ permutes lower sheets of the 
curve $\Gamma_3$;   $I_2$ and $I_3$ are complex conjugate. If $H > 0$, then $\tau_1$ 
leaves lower sheets invariant;  $I_2$ and $I_3$ are real.

\proclaim{Lemma  12} Let $p_n=0,\; n=1,\hdots, N$. Then the curve $\Gamma_N$ admits 
involution 
$$
\tau_-:\;\; (k,z)\rightarrow (-k,-z)
$$
The involution $\tau_{-}$ leaves the upper sheet invariant and 
$k^{(n)}_1=0$ for $N$ even. It can permute lower sheets and in this case 
$k_{\alpha}^{(n)}=k_{\tau_{-} \alpha}^{(n)}$ for $n$ odd and 
$k_{\alpha}^{(n)}=-k_{\tau_{-} \alpha}^{(n)} $ for $n$ even.

$\sigma=1$. All $k^{(0)}_{\alpha},\;\; \alpha =2,\hdots, N$ are pure imaginary.
\endproclaim
\demo\nofrills{Proof.\usualspace} Using the identity $\Phi(q,-z)=-\Phi(-q,z)$ we have 
$$
L(q,p=0,-z)+2(-k)=\[L(q,p=0,z)+2 k\]^{T} (-1)^N.
$$
Therefore $R(-k,-z)=(-1)^NR(k,z)$ and the existence of $\tau_-$ is proved. 

If a pair $(k,z)$ satisfies
$$
k={1\over z}k_{\alpha}^{(-1)} +k_{\alpha}^{(0)} + k_{\alpha}^{(1)}z+\hdots .
$$
Then
$$
-k={1\over -z}k_{\tau_-\alpha}^{(-1)} +k_{\tau_-\alpha}^{(0)} + 
k_{\tau_-\alpha}^{(1)}(-z)+\hdots .
$$
Therefore $k_{\alpha}^{(n)}=k_{\tau_-  \alpha}^{(n)}$ for n odd and 
$k_{\alpha}^{(n)}=-k_{\tau_{-} \alpha}^{(n)} $ for $n$ even.  Since $k_1^{(-1)}=N-1$ 
and $k_{\alpha}^{(0)}=-1,\; \alpha=2,\hdots, N$; $\tau_-$ leaves the upper sheet 
invariant. The involution $\tau_-$ can permute the lower sheets. It is demonstrated by 
the example for $N=3$.

$\sigma=1$. Skew-selfadjointness of the matrix $L^{(0)}$ is restored under the 
conditions of the Lemma.
\qed
\enddemo

\subhead  6. Appendix\endsubhead  
The Weierstrass $\sigma(z)$ has periods $2\omega$ and $2\omega'$ and is defined as 
$$
\sigma(z)=z\prod
 \{(1-{z\over  \boldsymbol\omega}) \exp({z\over \boldsymbol\omega }+{z^2\over 2
\boldsymbol\omega^2}) \},
$$
where $\boldsymbol\omega=2\omega n + 2\omega' n'$ and 
$\prod$ is taken with $n,n' \in \Z^1; \;
n^2+n^{\prime 2} >0$.  $\zeta(z)$ and $\wp(z)$ are difined similarly:
$$
\zeta(z)={d\over d\, z} \log \sigma(z), \quad \quad \wp(z)= - {d\over d\, z} \zeta(z).
$$
For more information see \cite{HC}.

{\bf Acknowledgments.} Finally, I would like to thank I.M. Krichever and 
H.P. McKean for numerous stimulating discussions.

\bye